\documentclass[aps,prd,twocolumn,groupedaddress,nofootinbib]{revtex4-2}

\usepackage{graphicx}
\usepackage{xcolor}
\usepackage{amsmath}
\usepackage{ulem}
\usepackage{hyperref}
\usepackage{comment}
\usepackage{xurl}
\usepackage{aas_macros}
\begin{document}

\title{Probing photon-ALP oscillations from the MAGIC observations of FSRQ QSO B1420+326}


\author{Bhanu Prakash Pant}
\email{pant.3@iitj.ac.in}
\affiliation{Department of Physics, Indian Institute of Technology Jodhpur, Karwar 342037, India.}


\begin{abstract}
At the beginning of 2020, MAGIC reported a very-high-energy (VHE) flaring activity from the FSRQ QSO B1420+326. It is now the fourth known most distant blazar ($z$=0.682) with an observed VHE gamma-ray emission. In this work, we investigate the effect of photon--axionlike particle (ALP) oscillations in the gamma-ray spectra measured by \textit{Fermi}-LAT and MAGIC around the flaring state. We set 95\% C.L. upper limit on the ALP parameters and obtain a constraint on the photon-ALP coupling constant $g_{a\gamma} < 2\times10^{-11}$ GeV$^{-1}$ for ALP masses $m_{a} \sim 10^{-10}-10^{-9}$ eV. Assuming the hadronic origin of VHE photons, we also estimate the expected neutrino flux from this source and the contribution to diffuse neutrino flux from QSO B1420+326-like FSRQs at sub-PeV energies. Furthermore, we study the implications of photon-ALP oscillations on the counterpart $\gamma$-rays of the sub-PeV neutrinos. Finally, we investigate a viable scenario of invisible neutrino decay to ALPs on the gamma-ray spectra and diffuse $\gamma$-ray flux at sub-PeV energies. Interestingly, we find that for the choice of neutrino decay lifetime $\tau_{2}/m_{2} = 10^3$ s eV$^{-1}$, the $\gamma$-ray flux has a good observational sensitivity towards LHAASO-KM2A.
\end{abstract}


\maketitle
\section{\label{sec:intro}Introduction}
Axionlike particles (ALPs) are ultralight pseudoscalar (spin 0) bosons proposed as an extension of physics beyond the Standard Model (BSM) \cite{svrcek2006axions,arvanitaki2010string} similar to QCD axions by Peccei and Quinn to solve the strong \textit{CP} problem \cite{peccei1977constraints, peccei1977cp}. They have weak coupling to Standard Model (SM) particles and are potential candidates for dark matter, and thus may account for its significant fraction in the Universe \cite{preskill1983cosmology,abbott1983cosmological,dine1983not,sikivie2010dark}. ALPs can couple to photons via coupling strength $g_{a\gamma}$ in the presence of an external electromagnetic field resulting in photon-ALP oscillations. In contrast to QCD axions, ALP mass $m_{a}$ and $g_{a\gamma}$ are treated as independent parameters.

Many searches have been performed to detect these ALPs exploiting photon-ALP oscillations. From the nondetection of these photons, several bounds have been placed by laboratory experiments \cite{zavattini2006, duffy2006, bregant2008, pugnat2008, ehret2010, ballou2015}. So far, the stringent bound on ALP parameters is given by CERN Axion Solar Telescope (CAST) \cite{zioutas1999decommissioned}, with $g_{a\gamma} <$ 6.6$\times$10$^{-11}$ GeV$^{-1}$ for $m_a < 0.02$ eV \cite{cast2017}. In the near future, experiments like Any Light Particle Search (ALPS) II \cite{bahare2013}, STAX \cite{capparelli2016}, International Axion Observatory (IAXO) \cite{armengaud2019}, and ABRACADABRA \cite{ouellet2019} will provide more stringent constraints on the ALP parameter space. 

Apart from the laboratory experiments, another promising avenue is to look at $\gamma$-rays originating from astrophysical sources. While propagating from higher-redshift sources, these VHE ($>$ 100 GeV) $\gamma$-rays suffer attenuation by extragalactic background light (EBL) or cosmic microwave background (CMB). Under the photon-ALP mixing, the transparency of these VHE photons increases drastically, leading to modulation in their observed $\gamma$-ray spectra. Detecting these VHE fluxes by $\gamma$-ray detectors may provide crucial hints on photon-ALP mixing. Many works have been performed by studying the $\gamma$-ray spectra of several Galactic and extragalactic sources \cite{hooper2007detecting, mena2011,meyer2013first, abramowski2013constraints, ajello2016search,  meyer2017fermi, liang2019constraints, calore2020, bi2021axion, caputo2021, guo2021implications, schiavone2021, li2021limits, fiorillo2022, mastrotaro2022, davies2023, pant2023}. Most noticeable is the recent observation of $\sim$18 TeV photons by Large High Altitude Air Shower Observatory (LHAASO) with the kilometer square area (KM2A) \cite{lhaasogcn} and an astonishing $\sim$251 TeV photon by Carpet-2 \cite{carpetgcn} from a long gamma-ray burst, GRB 221009A at redshift 0.1505. In a conventional scenario, such high-energy (HE) photons should be attenuated by EBL; therefore, some unconventional physics, e.g., photon-ALP oscillations, seems to be involved \cite{wang2023, galanti2023}.
 
In this work, we focus on the observations of VHE $\gamma$-ray spectra of QSO B1420+326, also known as OQ 334, by the Major Atmospheric Gamma Imaging Cherenkov Telescopes (MAGIC) \cite{acciari2021}. It is the fourth most distant blazar of redshift 0.682 with an observed VHE emission. It is classified as the flat-spectrum radio quasar (FSRQ) \cite{healey2007}. The source was repeatedly observed in the HE state from its first observation above 10 GeV by \textit{Fermi}-LAT \cite{ciprini2018, angioni2019, ciprini2020}. MAGIC performed follow-up observations and, at the beginning of January 2020, reported an enhanced activity from the source. The VHE emission detected was estimated to be about 15\% of the Crab Nebula flux above 100 GeV. Alerts have been sent to various observatories for follow-up observations from radio to VHE $\gamma$-rays \cite{dammando2020, ramazani2020, minev2020}. The first significant detection ($\sim$14.3$\sigma$) of VHE flare from QSO B1420+326 by MAGIC was achieved on January 20, 2020 in 1.6 h of exposure time. In this period, the flux reached $\sim$7.8$\times 10^{-11}$ cm$^{-2}$ s$^{-1}$ above 100 GeV. Further hints of significant excess were obtained in subsequent days after the VHE flare, namely post-flare, which lasted until February 1, 2020. The highest excess ($\sim$6.6$\sigma$) in the post-flare phase was obtained on January 31, 2020, with the longest exposure time of 2.5 h. Since FSRQs in a flaring state provide significant statistics to VHE $\gamma$-ray observatories, this makes them a good candidate source to study photon-ALP oscillations.  

This paper is structured as follows. In Sec. \ref{sec:alposc}, we briefly describe the photon-ALP mixing in an external magnetic field. Section \ref{sec:magenv} describes the various magnetic field environments considered in this work. In Sec. \ref{sec:fermianal}, we describe the Fermi-LAT analysis of QSO B1420+326. In Sec. \ref{sec:methodology}, we describe our data fitting methodology on the observed $\gamma$-ray spectra. In Sec. \ref{sec:resdis}, we discuss our constraints on the ALP parameters. We also give an estimate of the expected neutrino flux and the cumulative emission from QSO B1420+326-like sources at sub-PeV energies. We then discuss the implications of photon-ALP oscillations on the neutrino counterpart $\gamma$-rays and diffuse $\gamma$-ray flux. Finally, we discuss a viable scenario of invisible neutrino decay to ALPs and its implications on sub-PeV $\gamma$-ray spectra.

\section{\label{sec:alposc}photon-ALP oscillations}
The minimal interaction between photons and ALPs in the presence of an external magnetic field can be described by
\begin{equation}
    \mathcal{L}_{int} = \frac{-1}{4}g_{a\gamma} \,a\,F_{\mu \nu} \tilde{F}^{\mu \nu}=g_{a\gamma}\,a\,\textbf{E} \cdot \textbf{B},
\end{equation}
where $g_{a\gamma}$ is the coupling between photons and ALPs, $F_{\mu \nu}$ is the electromagnetic field tensor, $\tilde{F}^{\mu \nu}$ is the dual tensor, \textbf{E} is the electric field of the propagating photon beam, and \textbf{B} is the external magnetic field. 

Consider an initially polarized, monoenergetic beam of photons with energy $E$ propagating along the $\hat{\textbf{z}}$ direction. If the propagating medium is filled with a homogeneous external magnetic field \textbf{B} along the $\hat{\textbf{y}}$ axis, the equation of motion, in the limit $E \gg m_{a}$, is given by \cite{raffelt1988mixing}
\begin{equation}
\left(i\frac{d}{dz}+E+\mathcal{M}_{0}\right) \psi(z) = 0 \, .
\end{equation}
with $\psi(z) = \left(A_{x}(z), A_{y}(z), a(z)\right)^T $, where $A_{x}(z)$, $A_{y}(z)$, and $a(z)$ denote the photon amplitudes with transverse polarization states along the x and y axis, and amplitude associated with ALP field, respectively, while $\mathcal{M}_{0}$ represents the photon-ALP mixing matrix.

We can neglect the contribution of the QED vacuum polarization for weak magnetic fields. Furthermore, we can neglect the effect of Faraday rotation since we are considering the energy $E$ in the  VHE $\gamma$-rays regime. This leads to the simplification of the form of the mixing matrix 
\begin{equation}
\mathcal{M}_{0} = 
\begin{pmatrix}
\Delta^{xx} & 0 & 0\\
0 & \Delta^{yy} & \Delta^{y}_{a\gamma} \\
0 & \Delta^{y}_{a\gamma} & \Delta^{zz}_{a}
\end{pmatrix}
\, ,
\end{equation}
with $\Delta^{xx} = \Delta^{yy} = -\omega^{2}_{pl}/2E$, $\Delta^{zz}_{a} = -m^{2}_{a}/2E$, and $\Delta^{y}_{a\gamma} = g_{a\gamma\gamma}B_{y}/2$. Here, $\omega^{2}_{pl}$ is the plasma frequency resulting from the effective photon mass arising from the charge screening effect as the beam propagates through the cold plasma.

The transport matrix, $T(s) = T(s_{N}) \times T(s_{N-1}) \times ... \times T(s_{1})$, of the photon-ALP beam for the whole propagation length can be written by splitting it into \textit{N} sub-regions assuming a constant magnetic field in each region. The final photon survival probability in the photon-ALP system can be written as
\begin{equation} 
\label{galp-prob}
    P_{\gamma\gamma} = \text{Tr}\left[(\rho_{11}+\rho_{22})T(s)\rho(0)T^{\dagger}(s)\right],
\end{equation}
where $\rho(0) = \frac{1}{2} \text{diag}(1, 1, 0)$ is the initial polarization of the beam, $\rho_{11} = \text{diag}(1, 0, 0)$ and $\rho_{22} = \text{diag}(0, 1, 0)$ denotes the polarization along the \textit{x} and \textit{y} axis, respectively.

In the strong-mixing regime, $E_{crit}\leq E \leq E_{max}$, photon-ALP oscillations probability becomes independent of energy. It becomes maximal with $ E_{crit} = |m_{a}^{2}-\omega_{pl}^{2}|/2g_{a\gamma}B_{T}$ and $E_{max} = 90\pi B_{cr}^{2}g_{a\gamma}/ 7\alpha B_{T}$ where, $m_{a}$ is the mass of the ALP field, $\omega_{pl} = 3.69 \times 10^{-11} \left(n_{e}/cm^{-3}\right)^{1/2}$ is the plasma frequency, $B_{cr} = m_{e}^2/|e| = 4.4 \times 10^{13}$ G is the critical magnetic field, $\alpha$ is the fine-structure constant, and $B_{T}$ is the transverse component of the external magnetic field.

\section{\label{sec:magenv} Magnetic field environments}
In this section, we summarize the various magnetic field environments considered in our calculation where the photon-ALP conversion of the beam can take place.

\subsection{\label{subsec:bjmf}Blazar jet region}
First, we consider the photon-ALP oscillations in the blazar jet magnetic field (BJMF) at the source. The BJMF can be modeled with a toroidal ($B \propto r^{-1}$) and a  poloidal ($B \propto r^{-2}$) components. In this work, we consider only the toroidal component since the latter diminishes at large distances from the black hole center. The magnetic field strength of the BJMF can be written as\cite{begelman1984theory,ghisellini2009canonical}
\begin{equation}
    B_{jet}(r) = B_{jet}(0)\left(\frac{r}{r_{VHE}}\right)^{-1},
\end{equation}
where $r_{VHE}$ is the distance of the VHE $\gamma$-ray emission site to the central black hole and $B_{jet}(0)$ is the magnetic field strength at $r_{VHE}$. We assume the magnetic field strength is negligible for the jet region $>$1 kPc.

We consider the electron density profile following a power law given as \cite{o2009magnetic}
\begin{equation}
    n_{el}(r) = n_{el}(0)\left(\frac{r}{r_{VHE}}\right)^{\beta},
\end{equation}
where $n_{el}(0)$ is the electron density at $r_{VHE}$. Here, we consider $\beta=2$ assuming equipartition between the magnetic field and electrons. A more realistic model accounting for the fact that electron distribution is nonthermal in a relativistic AGN jet is provided in Ref. \cite{davies2022relevance}. 

It is to be noted that the above equations hold in the comoving jet frame with photon energy $E^{'}$ related to the energy $E$ in the lab frame by $E^{'} = E/\delta$, where $\delta = \left[\Gamma_{L}(1-\beta^{2}cos\theta_{obs})\right]^{-1}$ is the Doppler factor with $\Gamma_{L}$ and $\beta$ as the bulk Lorentz and beta factor, respectively, and $\theta_{obs}$ is the angle between the jet axis and the line of sight.

Table \ref{table:source_params} lists the BJMF model parameters values for QSO B1420+326 used in our analysis and taken from Ref. \cite{acciari2021}.
\begin{table}
\begin{tabular}{c c c}
\hline
 \textbf{Parameter name} & \textbf{VHE flare} & \textbf{Post-Flare}  \\ 
 \hline
 R.A.(J2000) & 14 22 30.38 (hh mm ss) & "\\
 Dec.(J2000) & +32 23 10.44 (dd mm ss) & "\\ 
 z & 0.682 & " \\
$\theta_{view}$ [deg] & 0.8 & " \\ 
 $\delta$ & 40 & " \\
 $\Gamma$ & 40 & "  \\
 $B^{Jet}_{0}$ [G] & 0.83 & 0.55  \\
 $u^{'}_{e}$ [erg.cm$^{-3}$] & 17.3 $\times$ 10$^{-3}$ &  19.2 $\times$ 10$^{-3}$ \\
 $R^{'}_{blob}$ [10$^{16}$ cm] & 3.08 & "\\
 $\eta$ & -1 & " \\
 $\xi$ & -2 & " \\
 $\gamma_{e,min}$ & 10 & " \\
 $\gamma_{e,max}$ & 23700 & 27300 \\
 \hline
\end{tabular}
\caption{\label{table:source_params}Summary of the BJMF model parameters in the VHE flare and post-flare states taken from Ref. \cite{acciari2021}.}
\end{table}

\subsection{\label{subsec:icmf} Intracluster region}
After leaving the jet, the photon-ALP beam may enter a rich cluster environment where the blazar is located. The strength of the turbulent magnetic field is $\sim$1 $\mu$G \cite{carilli2002cluster,govoni2004magnetic,subramanian2006evolving}, and the photon-ALP effect could be significant \cite{meyer2014detecting}. The intracluster magnetic field (ICMF) can be modeled as
\begin{equation}
    B_{ICMF}(r) = B_{ICMF}(0) \left(\frac{n_{el}(r)}{n_{el}(r_{0})}\right)^{\xi},
\end{equation}
where $B_{ICMF}(0)$ and $n_{el}(r_{0})$ are the magnetic field strength and electron density at the cluster center, respectively, $\xi$ ranges from $0.5-1$, and $n_{el}(r)$ is the electron density distribution given by
\begin{equation}
    n_{el}(r) = n_{ICMF}(0) \left(1 + \frac{r}{r_{core}}\right)^{\eta},
\end{equation}
with  $\eta=-1$ and $r_{core}$ as the core radius. The typical values of $B_{ICMF}(0)$, $n_{ICMF}(0)$, and $r_{core}$ are of the order of $\sim$1 $\mu$G, $\sim$10$^{-3}$ cm$^{-3}$ and $\sim$100 kpc, respectively.

Since there is no evidence that QSO B1420+326 is located in a rich cluster environment, we neglect the photon-ALP oscillations in this region.

\subsection{\label{subsec:egmf}Extragalactic region}
The cosmological scale of the extragalactic region is $\sim$$\mathcal{O}$(1) Mpc with $\sim$$\mathcal{O}$(1) nG of magnetic field strength \cite{ade2016planck, pshirkov2016new}. Therefore, the extragalactic magnetic field is too feeble to produce significant photon-ALP conversions and can be neglected. We consider only the absorption effect due to EBL/CMB with the optical depth \cite{belikov2011no} 
\begin{eqnarray}
    \tau (E_{\gamma},z) = c\,\int_{0}^{z} \frac{dz}{(1+z)H(z)} \, \int_{E_{th}}^{\infty} d\epsilon \, \frac{dn(z)}{d\epsilon} \, \nonumber \\ \times \, \tilde\sigma_{\gamma\gamma}(E_{\gamma}, \epsilon, z)\,,
\end{eqnarray}
where $ E_{th} = 2(m_{e}c^{2})^{2}/E_{\gamma}(1-cos\theta)$ is the threshold energy for pair-production with angle $\theta$ between the projectile and target photons of energy $E_{\gamma}$ and $\epsilon$, respectively, $z$ is the redshift of the source, $H(z)$ is the Hubble expansion rate, $dn(z)/d\epsilon$ is the proper number density of the target photons, and $\tilde\sigma_{\gamma\gamma}$ is the integral pair production cross section. Several EBL models are proposed in the literature \cite{franceschini2008extragalactic, kneiske2010lower, Finke2010, dominguez2011extragalactic, gilmore2012semi, franceschini2017extragalactic, saldana2021observational}, we consider the EBL model by Dom\'inguez \textit{et al.} \cite{dominguez2011extragalactic} in this work. 

\subsection{\label{subsec:gmf}Galactic region}
In the past few years, the knowledge of the magnetic field in the Milky Way region has been significantly improved. It is now known that the strength of the Galactic magnetic field (GMF) is of the order of $\sim$$ \mathcal{O}$($\mu$G) and comprises a regular and a turbulent component. The coherence length of the turbulent component is smaller than the photon-ALP oscillation length. Therefore, we consider only the regular component in this study.

In this work, we consider the GMF model by Jansson and Farrar \cite{jansson2012new}. In addition to the disk component, this model assumes a halo component parallel to the galactic plane and a poloidal component at the galactic center. In the updated version of this model \cite{Jansson_2012}, the data from Planck satellite \cite{adam2016planck} about the thermal electron distribution is considered. 

\section{\label{sec:fermianal}Fermi-LAT analysis of QSO B1420+326}
In the Fermi-LAT Fourth Source Catalog (4FGL) \cite{abdollahi2020}, the source QSO B1420+326 is associated with gamma-ray source '4FGL J1422.3+3223' with flux above 100 MeV. We perform Fermi-LAT data analysis for two phases, namely:
\begin{enumerate}
    \item \textbf{VHE flare :} January 20, 2020 (MJD 58868.3) to January 22, 2020 (MJD 58870.3).
    \item \textbf{Post flare :} January 22, 2020 (MJD 58873.5) to February 01, 2020 (MJD 58880.5).
\end{enumerate}
We use Fermi-LAT Pass 8 processed data from Fermi Science Data Center (FSDC) \footnote{\label{foot:fermidata}\url{https://fermi.gsfc.nasa.gov/ssc/data/access/}} for the above-mentioned periods and adopt the P8R3\_SOURCE\_V2 for instrument response functions (IRFs). We select the SOURCE class (evclass=128 and evtype=3) with 10$^\circ$ region of interest (ROI) centered on the target source. The data are binned into 0.1$^\circ$ angular bins and 8 bins per decade in the energy range of 100 MeV to 300 GeV. We apply zenith angle $<$ 90$^\circ$ cut to eliminate events from the Earth limb and consider all the 4FGL sources around 15$^\circ$ from the ROI center as background sources. We use preprocessed templates of Galactic diffuse emission, \texttt{gll\_em\_v08.fits}, and the extragalactic isotropic diffuse emission, \texttt{iso\_P8R3\_SOURCE\_V2.fits}. We utilized the standard Python-based package Fermipy\footnote{\url{https://fermipy.readthedocs.io/en/latest/index.html}} \cite{2017ICRC...35..824W} for the likelihood analysis and the spectral energy distribution (SED).

\section{\label{sec:methodology}Methodology}
We consider the Fermi-LAT and MAGIC \cite{acciari2021} data points for the two phases and fit them under the null hypothesis. We take the intrinsic spectrum of QSO B1420+326 to be an exponential cutoff power law (EPWL)
\begin{equation}
   \Phi_{int}(E) = N_0 \left(\frac{E}{E_0}\right)^{-\alpha} \text{exp}\left(\frac{-E}{E_{cut}}\right) , \label{eq:intspect}
\end{equation}
where the reference energy, $E_0$, is kept fixed at 1 GeV and $N_0$, $\alpha$, and $E_{cut}$ are treated as free parameters. Table \ref{table:bestfitparams} summarizes the best-fit spectral parameters obtained along with 1$\sigma$ uncertainty. It is to be noted that we also test other forms of the intrinsic spectrum and find that the EPWL best fits the \textit{Fermi}-LAT and MAGIC data points. In Ref. \cite{li2022relevance}, it is shown that the choice of the intrinsic spectrum has no significant effect in constraining the ALP parameters.
\begin{table}
\begin{ruledtabular}
\caption{\label{table:bestfitparams}Summary of the best-fit spectral parameters with 1$\sigma$ uncertainty shown in the bracket.}
\begin{tabular}{c c c c c }
 Phase & $N_0$ (x10$^{-10}$) & $\alpha$ & $E_{cutoff}$ \\ 
  & [MeV$^{-1}$cm$^{-2}$s$^{-1}$] & & [GeV]\\
 \hline
VHE flare & 1.86(0.19) & 1.87(0.08) & 50.90(17.37)\\
Post-flare & 1.33(0.11) & 1.99(0.05) & 46.49(8.06)\\
\end{tabular}
\end{ruledtabular}
\end{table}

Under the assumption of photon-ALP oscillations, the survival probability of photons gets modulated, and the expected gamma-ray spectrum is given by
\begin{equation}
    \phi_{alp}(E) = \Phi_{int}(E)\cdot \mathcal{P}_{alp}(E)
\end{equation}
where $\mathcal{P}_{alp}$ is the survival probability of photons under the ALP hypothesis. We used publicly available \texttt{gammaALPs}\footnote{\url{https://gammaalps.readthedocs.io/en/latest/index.html}} \cite{Meyer:202199} package to calculate the photon-ALP conversion probability in the magnetic field environments discussed in Sec. \ref{sec:magenv}.

The best-fit ALP paramaters, $m_a$ and $g_{a\gamma}$, are obtained by minimizing the $\chi^2$ function
\begin{equation}
    \chi^2 = \sum_{i=1}^{N}\left(\frac{\Psi^{obs}_{i}-\phi^{exp}_{i}}{\sigma_{i}}\right)^2 \, ,
\end{equation}
where $\Psi^{obs}$ is the observed and $\phi^{exp}$ is the expected gamma-ray flux, with $\sigma$ being the corresponding uncertainty in the data.

\section{\label{sec:resdis}Results and Discussions}
\subsection{\label{subsec:alpconst}Constraints on ALP parameters}
\begin{figure*}
    \centering
    \includegraphics[width=\textwidth]{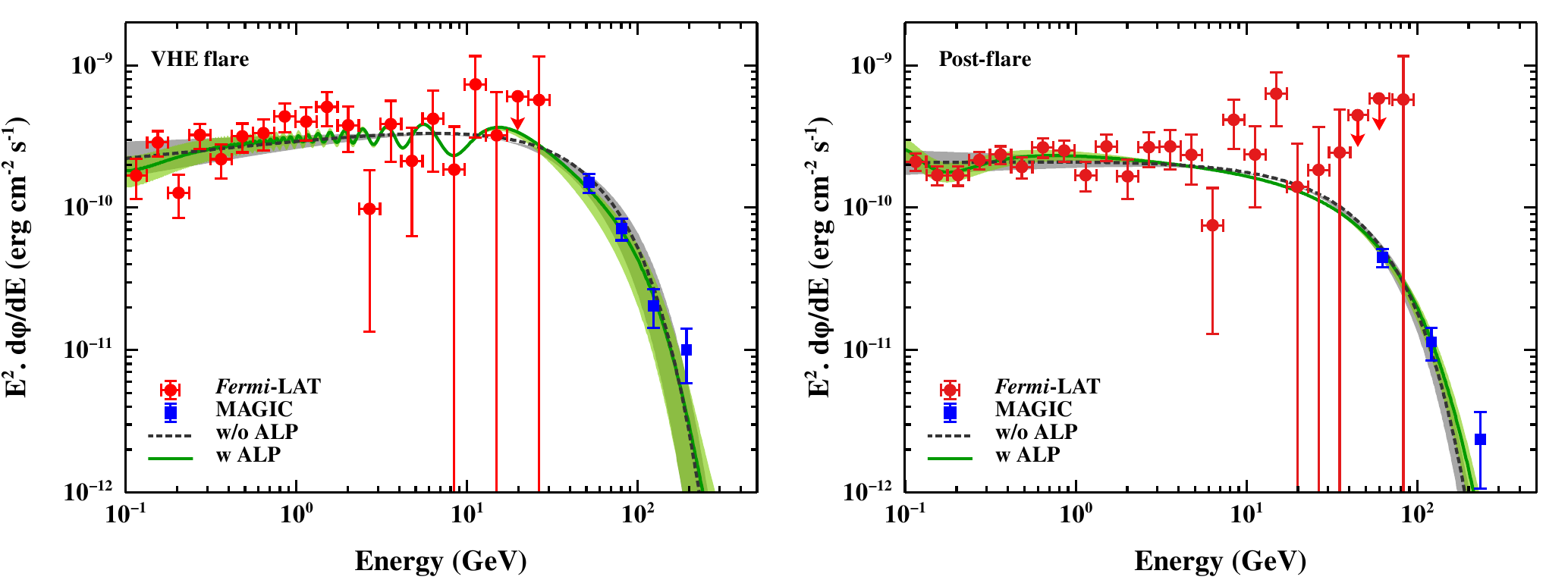}
    \caption{\label{fig:gaspect}Best-fit gamma-ray spectra of QSO 1420+326 for VHE flare (left) and post-flare(right). The dotted black and the solid green curves represent the spectra under the null and ALP hypotheses along with their 1$\sigma$ uncertainty band in light grey and light green colors. The best-fit ALP parameters $(m_{neV} = 3.68, g_{11}=5.30)$ and $(m_{neV} = 0.40, g_{11}=3.86)$ are used for the VHE flare and post-flare phase, respectively. The red circular and the blue square markers are the experimental data points from Fermi-LAT (See footnote \ref{foot:fermidata}) and MAGIC \cite{acciari2021}.}
\end{figure*}

\begin{figure*}
    \centering
    \includegraphics[width=\textwidth]{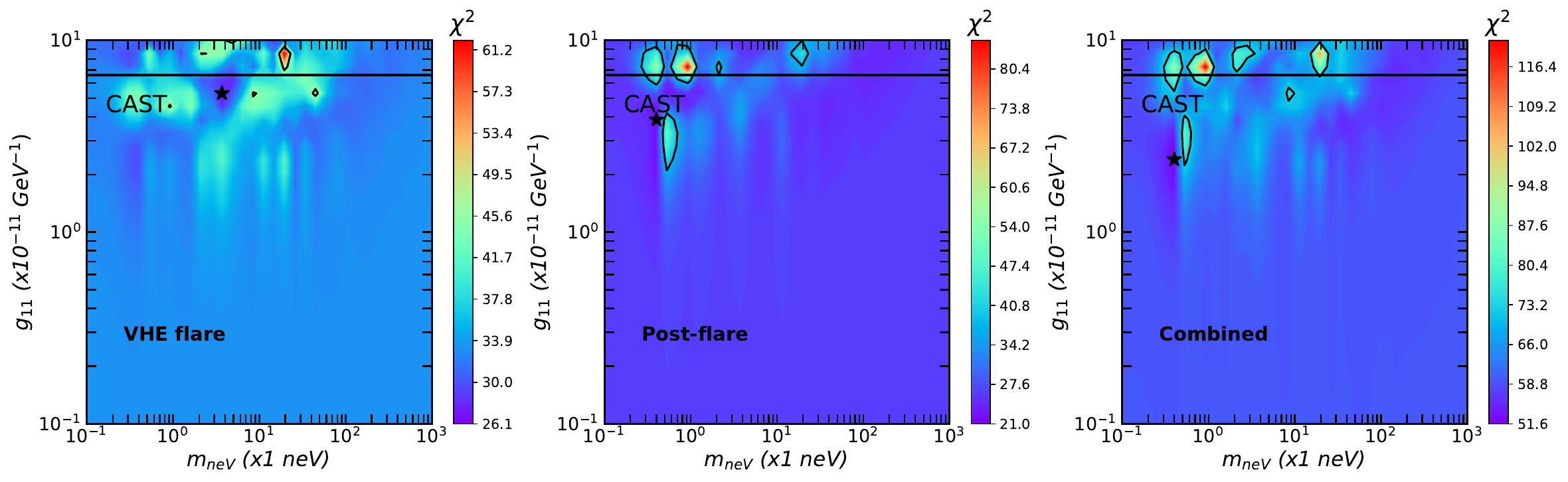}   \caption{\label{fig:chi2dist}Distribution of $\chi^{2}_{ALP}$ in the $m_{neV}$-$g_{11}$ parameter space for all two phases. The "$\star$" symbol in black represents the best-fit parameter point. The black contours represent the excluded parameter space at 95$\%$ C.L. in all two and the combined phases. The black horizontal line represents the upper limit set by the CAST experiment of g$_{a\gamma} <$ 6.6$\times$10$^{-11}$ GeV$^{-1}$ \cite{cast2017}.}
\end{figure*}

Using the methodology outlined in the previous section, we obtain the best-fit ALP parameters for each phase as summarized in Table \ref{table:bestfitchi2}.
\begin{table}[b]
\begin{ruledtabular}
\caption{\label{table:bestfitchi2}Summary of the best-fit $\chi^{2}$ values and ALP parameters under the null and ALP hypotheses.}
\begin{tabular}{c c c c c c}
 Phase & \textbf{$\chi^{2}_{w/o ALP}$} & \textbf{$\chi^{2}_{ALP}$} & $m_{neV}$ & $g_{11}$ & $\Delta \chi^{2}$   \\ 
 \hline
 VHE flare & 33.07 & 26.31 & 3.68 & 5.30 & 19.11\\
Post-flare & 26.34 & 21.33 & 0.40 & 3.86 & 15.65\\
\end{tabular}
\end{ruledtabular}
\end{table}
In Fig. \ref{fig:gaspect}, we show the best-fit $\gamma$-ray spectra under the null and ALP hypotheses. We use the best-fit ALP parameters $(m_{neV} = 3.68, g_{11}=5.30)$ and $(m_{neV} = 0.40, g_{11}=3.86)$ for the VHE flare and post-flare phase, respectively.
 The $\chi^2_{ALP}$ distribution in the $m_{neV}-g_{11}$ parameter space is shown in Fig. \ref{fig:chi2dist}. Here, we adopted the notations $m_{neV} \equiv m_{a}/1$ neV and $g_{11} \equiv g_{a\gamma}/10^{-11}$ GeV$^{-1}$.

\begin{figure*}
    \centering
    \includegraphics[width=\textwidth]{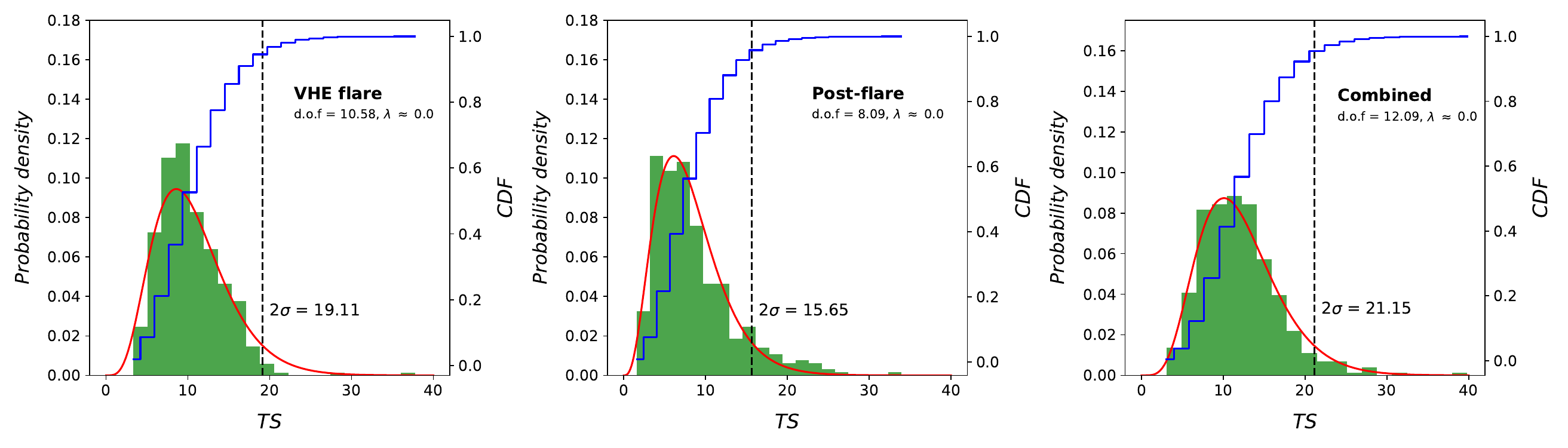}
    \caption{\label{fig:tsall}\textit{TS} distribution of VHE flare (left), post-flare (center), and the combined (right) phases of QSO B1420+326. The red curves show the fitted noncentral $\chi^{2}$ distributions. The blue lines show the cumulative density function (CDF) of the \textit{TS} distributions.}
\end{figure*}

In order to put constraints on ALP parameters, we determined $\chi^2_{thr} = \chi^2_{min} + \Delta\chi^2$ to exclude the region in each phase at a certain C.L. limit. Here, $\chi^{2}_{min}$ is the minimum $\chi^{2}$ value obtained in the  $m_{neV}-g_{11}$ plane and $\Delta\chi^2$ corresponds to a particular C.L. derived through Monte Carlo simulations. We perform 400 simulations for each phase, generating pseudodata by Gaussian samplings as in Ref. \cite{liang2019constraints}. For each set of pseudodata, we calculate the best-fit $\chi^2$ for both the null and ALP hypotheses as described in Sec.\ref{sec:methodology}. We calculate the test statistics, $TS = \chi^2_{null} - \chi^2_{ALP}$, which follows a noncentral $\chi^2$ distribution as shown in Fig. \ref{fig:tsall}. The $\Delta\chi^2$ values obtained by fitting these distributions in each phase are listed in Table \ref{table:bestfitchi2}. The black contours in Fig. \ref{fig:chi2dist} represent the excluded parameter space at 95\% C.L.

\begin{figure}
    \centering
    \includegraphics[width=\linewidth]{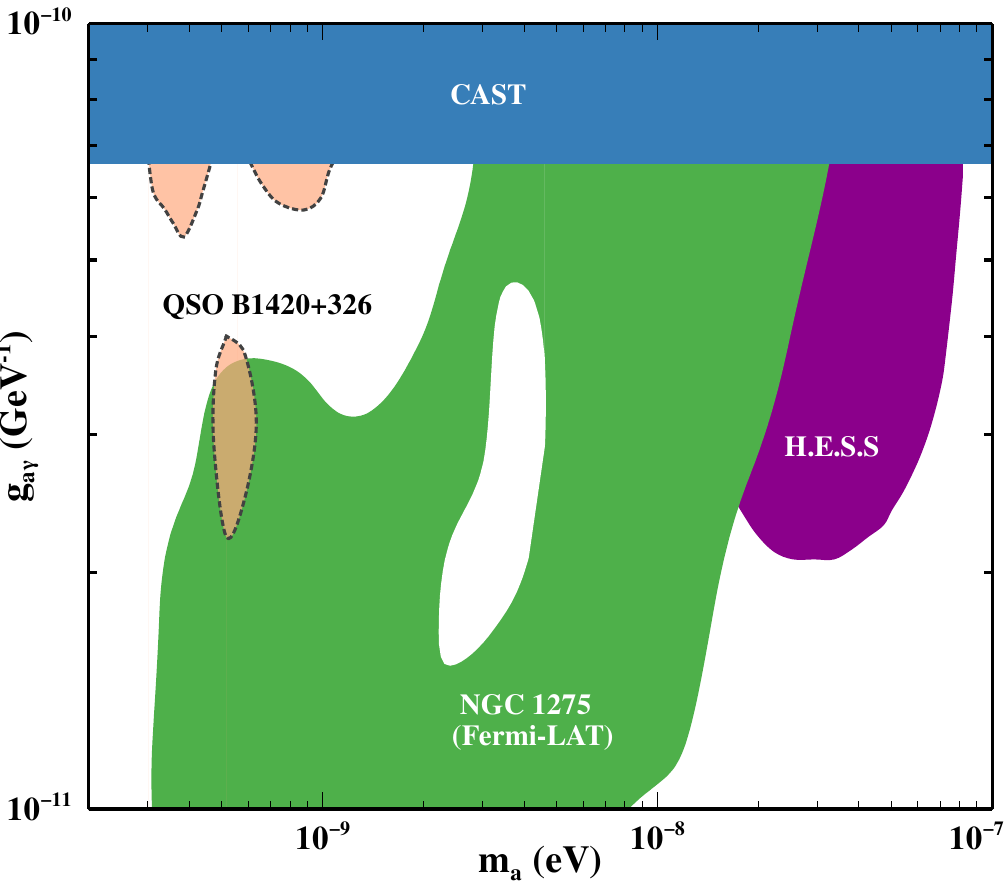}
\caption{\label{fig:alplimits}Expanded view of the exclusion region at 95\% C.L. for QSO B1420+326. We also show the constraints set by CAST \cite{cast2017}, NGC 1275 \cite{ajello2016search}, and H.E.S.S. \cite{abramowski2013constraints} for comparison.}
\end{figure}

We find weaker constraints as compared to CAST in the case of VHE flare. For the post-flare phase, a narrow region with $2\times10^{-11}$ GeV$^{-1}\le g_{a\gamma}\le 4\times$10$^{-11}$ GeV$^{-1}$ for $\sim 10^{-10}$ eV $\le m_{a} \le 10^{-9}$ eV is excluded. The combined constraint and some recent constraints in this ALP mass range are shown in Fig. \ref{fig:alplimits}.

\subsection{\label{subsec:expnuflux}Expected neutrino flux from QSO B1420+326}
FSRQs emit across the entire electromagnetic spectrum and can make up, among others, for some of the brightest $\gamma$-ray sources in the sky. It is usually believed that the low-energy emission is due to synchrotron photons by relativistic electrons in the plasma. In contrast, the high-energy emission is due to inverse Compton (IC) emission by upscattering either their own synchrotron photons or other external photon fields.
Another possible mechanism for producing VHE photons is through the hadronic channel, either $p-\gamma$ or $p-p$, leading to the production of neutral pions ($\pi^{0})$. These neutral pions then decay to VHE photons, which may be detected by ground-based detectors like MAGIC, High Energy Stereoscopic System (H.E.S.S), Cherenkov Telescope Array (CTA), and LHAASO. In addition to the neutral pions, charged pions are also produced, which eventually decay to neutrinos. The detection of $\sim$290 TeV neutrino from TXS 0506+056 blazar \cite{icecube2018multimessenger,icecube2018neutrino}  and neutrino emission from the active galactic nuclei (AGN) NGC 1068 \cite{IC2022ngc1068} by IceCube firmly establishes the hadronic models.

\begin{figure*}
    \centering
    \includegraphics[width=\textwidth]{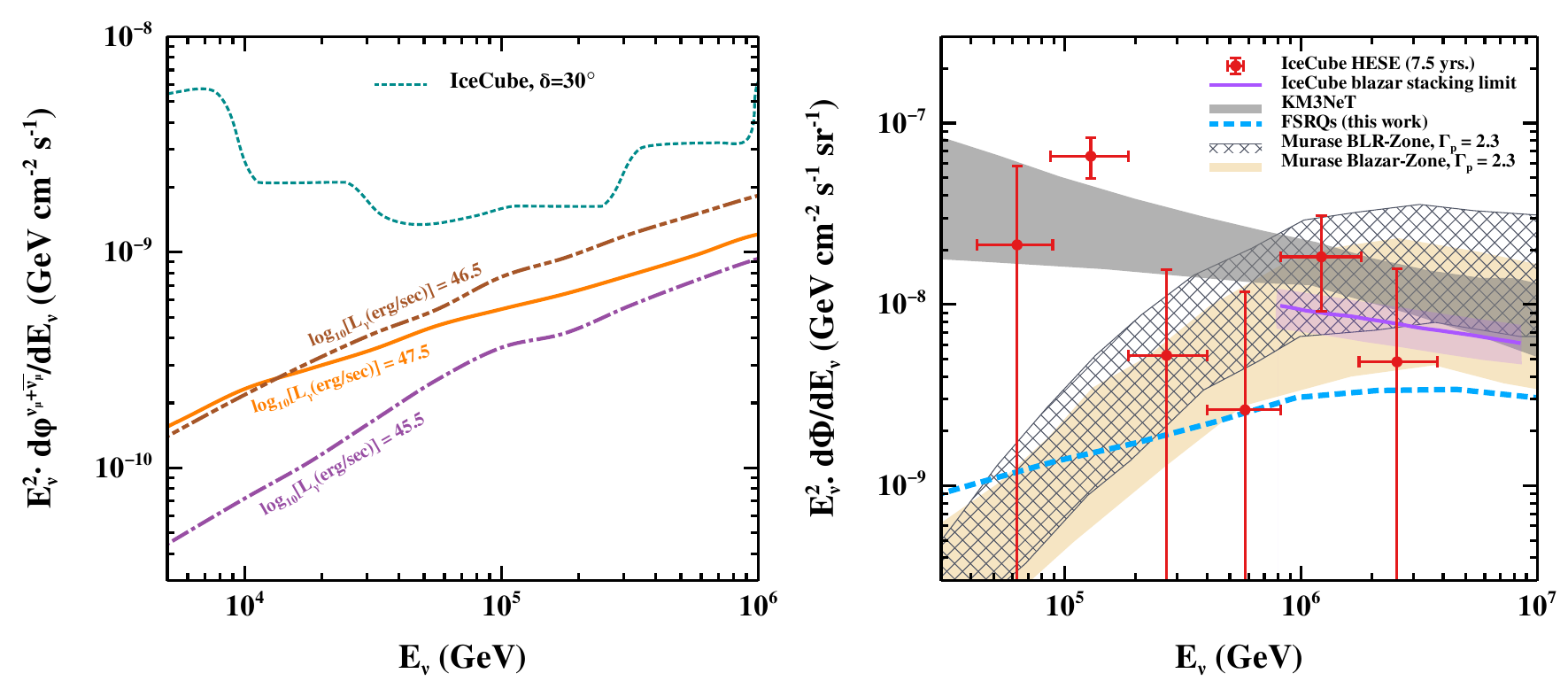}\caption{\label{fig:expnuflux}Left: expected muon neutrino plus antineutrino flux from QSO B1426+326 at three benchmark values of $L_{\gamma}$. The IceCube differential sensitivity \cite{aartsen2019search} for point-like sources is shown by a dotted curve at the nearest declination of the source. Right: diffuse neutrino flux (blue dashed curve) by convolving the single point-source flux of QSO B1420+326-like sources along with IceCube HESE events (7.5 yrs.) \cite{IC_HESE2021}, IceCube blazar stacking limit \cite{IC_stack2017}, KM3NeT sensitivity for diffuse flux \cite{caiffi2021}, and neutrino flux estimated from the inner jet model by Murase et al. \cite{murase2014}}
\end{figure*}

In this section, we estimate the expected neutrino flux at sub-PeV energies from QSO B1420+326, assuming VHE photons observed by MAGIC originated from neutral pion decay.
The flux of astrophysical neutrinos, $\phi_{src}$, at Earth from a single FSRQ can be written as \cite{palladino2019}
\begin{equation}
    \frac{d\phi_{src}}{dE_{\nu}}(E_{\nu},L_{\gamma}, z, \eta(L_{\gamma})) = \frac{1}{4 \pi d(z)^2}\left[\frac{1}{E_{\nu}} \frac{dL_{\nu}}{dE_{\nu}}\right] \times \eta(L_{\gamma}) \label{eq:nusrc}
\end{equation}
where $d(z)$ is the comoving distance, $dL_{\nu}/E_{\nu}dE_{\nu}$ is the neutrino luminosity spectra taken from Fig. 2 of Ref. \cite{palladino2019}, and $\eta(L_{\gamma}) = L_{CR}/L_{\gamma}$ is the baryonic loading, with $L_{CR}$ and  $L_{\gamma}$ as the luminosity of the injected CRs and the $\gamma$-ray luminosity of the source, respectively. Here, the baryonic loading is considered to evolve with $L_\gamma$ as a continuous function as in Ref. \cite{palladino2019}. Since the gamma luminosity for QSO 1420+326 is not yet constrained, we choose three benchmark values of 10$^{45.5}$ erg/sec, 10$^{46.5}$ erg/sec, and 10$^{47.5}$ erg/sec for $L_{\gamma}$, to calculate the neutrino flux. In the left panel of Fig. \ref{fig:expnuflux}, we show the expected sub-PeV neutrino flux along with the IceCube sensitivity \cite{aartsen2019search} for point sources at the nearest declination of QSO 1420+326. We find that for all three $\gamma$ luminosities, the neutrino flux has weak observational sensitivity towards the IceCube detector.

We also calculate the diffuse neutrino flux from FSRQs convolving the single point-source flux of QSO B1420+326-like sources with the source distribution over $L_{\gamma}$ and $z$ using 
\begin{eqnarray}
    \Phi_{diff}(E_{\nu}) &=& \int_{\Gamma_{min}}^{\Gamma_{max}} {\frac{dN}{d\Gamma} \, d\Gamma }\int_{z_{min}}^{z_{max}} \frac{d^{2}V}{dz d\Omega} \, dz \int_{L_{\gamma}^{min}}^{L_{\gamma}^{max}} \, dL_{\gamma} \, \nonumber \\ & &\times \,  \rho(L_{\gamma},z) .\, \frac{d\phi_{src}}{dE_{\nu}}(E_{\nu}, L_{\gamma}, z, \eta(L_{\gamma})) \,, \label{eq:diffnuflux}
\end{eqnarray}
where $dN/d\Gamma$ is the intrinsic photon index distribution which is assumed to be a Gaussian, $d^{2}V/dz d\Omega$ is the comoving volume element per unit redshift per unit solid angle, $d\phi_{src}/dE_{\nu}$ is the neutrino spectra, here taken as obtained for QSO B1420+326 for $L_{\gamma} = 10^{47.5}$ erg/sec, and $\rho(L_{\gamma},z)$ is the gamma-ray luminosity function (GLF). We consider here the luminosity-dependent density evolution (LDDE) of the GLF with parametrization as given in Ref. \cite{ajello2012}. The limits of integration are $\Gamma_{min}=1.8$, $\Gamma_{max}=3.0$, $z_{min}=0.01$, $z_{max}=3$, $L_{\gamma}^{min}=10^{46}$ erg/sec, and $L_{\gamma}^{max}=10^{51}$ erg/sec.

The resulting diffuse neutrino flux is shown in the right panel of Fig.\ref{fig:expnuflux}. For comparison, we also show the IceCube high-energy starting events (HESE) \cite{IC_HESE2021}, IceCube blazar stacking limit \cite{IC_stack2017}, Cubic Kilometre Neutrino Telescope (KM3NeT) sensitivity for diffuse flux \cite{caiffi2021}, and neutrino flux estimated from the inner jet model by Murase \textit{et al.} \cite{murase2014}. We find that FSRQs can provide sub-dominant contribution to the extragalactic diffuse neutrino flux at sub-PeV energies.

\subsection{\label{susec:subpevgamma}Counterpart $\gamma$-rays at sub-PeV energies}
In this section, we estimate the residual gamma-ray flux under the ALP hypothesis as a counterpart of sub-PeV neutrinos. We obtain the gamma-rays flux at the source using the relation \cite{PhysRevD.78.034013} $E^{2}_{\gamma}\cdot dN_{\gamma}/dE_{\gamma}= (2/3) E^{2}_{\nu}\cdot dN_{\nu}/dE_{\nu}$, where $E_{\gamma} = 2 E_{\nu}$ as a consequence of $\pi^{0}$ decay. These VHE photons undergo attenuation by synchrotron and synchrotron self-Compton (SSC) photons due to relativistic electrons inside the blob. The escape fraction of these VHE photons of energy $\epsilon^{\prime}_{\gamma}$ (in $m_{e}c^{2}$) in the jet frame is given by
\begin{equation}
  \mathcal{P}^{esc}_{\gamma\gamma}(\epsilon^{\prime}_{\gamma}) = \frac{1-\exp{(-\tau_{\gamma\gamma}(\epsilon^{\prime}_{\gamma})})}{\tau_{\gamma\gamma}(\epsilon^{\prime}_{\gamma})} \, , \label{eq:escfrac}
\end{equation}
where $\tau_{\gamma\gamma}(\epsilon^{\prime}_{\gamma})$ is the optical depth of this interaction \cite{PhysRevD.99.103006}
\begin{equation}
  \tau_{\gamma\gamma}(\epsilon^{\prime}_{\gamma}) = R^{\prime}_{blob} \int \sigma_{\gamma\gamma}(\epsilon^{\prime}_{\gamma}, \epsilon^{\prime}_{k}) \, n^{\prime}_{k}(\epsilon^{\prime}_{k}) \, d\epsilon^{\prime}_{k} \, .
\end{equation}
where $R^{'}_{blob}$ is the blob radius, $\sigma_{\gamma\gamma}$ is the pair production cross section, $n^{\prime}_{k}(\epsilon^{\prime}_{k})$ is the number density of the ambient photons of energy $\epsilon^{\prime}_{k}$ (in $m_{e}c^{2}$) in the jet frame.

As these survived photons propagate over cosmic distances, they again interact with the CMB photons, initiating electromagnetic cascades, and gets exhausted. Under the ALP hypothesis, these photons may convert into ALPs that can propagate unimpeded. Upon entering the Galactic magnetic field, these ALPs may backconvert into photons and may be observed as a residual flux at sub-PeV energies.

\begin{figure*}
    \centering
    \includegraphics[width=\textwidth]{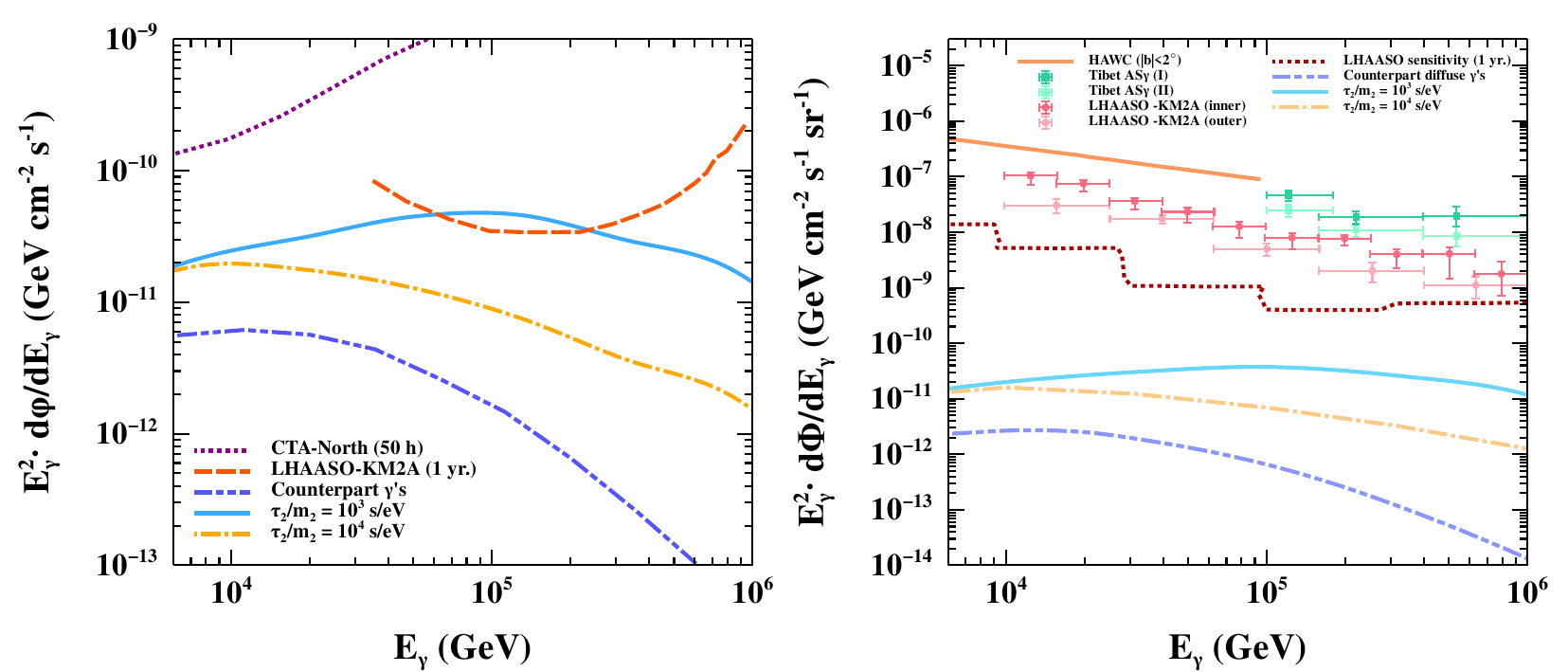}\caption{\label{fig:expgaflux}Left: counterpart $\gamma$-ray flux from QSO B1426+326 for $L_{\gamma} = 10^{47.5}$ erg/sec. The differential sensitivity for Crab-like point gamma-ray sources by CTA-North (dotted) for 50 h of exposure and LHAASO-KM2A \cite{Vernetto_2016} (dashed) for 1 year of exposure is also shown. Right: counterpart diffuse $\gamma$-ray flux from FSRQs (dash-dot-dotted curve) along with Galactic diffuse $\gamma$-ray flux measured by HAWC \cite{alfaro2023galactic}, Tibet-AS$\gamma$ \cite{amenomori2021galactic}, and LHAASO-KM2A \cite{cao2023galactic}. The dotted brown curve is the LHAASO 1 yr. sensitivity to Galactic diffuse $\gamma$-ray flux \cite{neronov2020}. The solid blue and dot-dash yellow lines in both panels correspond to $\gamma$-ray flux from ALPs originated from invisible neutrino decay for two benchmark values of $\tau_{2}/m_{2}$.}
\end{figure*}

In the left panel of Fig. \ref{fig:expgaflux}, we show the counterpart $\gamma$-rays corresponding to the neutrino flux obtained for $L_{\gamma} = 10^{47.5}$ erg/sec. For comparison, we also show the CTA-North \footnote{\url{https://www.cta-observatory.org/science/ctao-performance}} and LHAASO-KM2A \cite{Vernetto_2016} differential sensitivity for Crab-like point gamma-ray sources. We find that the counterpart sub-PeV $\gamma$-rays under the ALP hypothesis have a weak sensitivity towards both detectors. 

We also estimate the diffuse $\gamma$-ray flux from FSRQs in analogy with the diffuse neutrino flux as in Eq. \ref{eq:diffnuflux}, which is shown in the right panel of Fig. \ref{fig:expgaflux}. For comparison, some recent observations of the Galactic diffuse $\gamma$-ray flux by High-Altitude Water Cherenkov observatory (HAWC) \cite{alfaro2023galactic}, Tibet-AS$\gamma$ \cite{amenomori2021galactic}, and LHAASO-KM2A \cite{cao2023galactic} are also shown.

\subsection{\label{subsec:invnudec}Implications of invisible neutrino decay on the sub-PeV $\gamma$-ray spectra}
In the SM of elementary particles, neutrinos were long believed to be massless. Over the past several decades, experimental evidence established the nonzero mass of neutrinos. They are now known to have three discrete tiny masses where a neutrino of a specific flavor is a superposition of these three mass eigenstates. In BSM scenarios, the heavier neutrinos could decay into lighter ones \cite{bahcall1972}, \footnote{In this work, we neglect the contribution from $\nu_{3}$ decay, i.e., $\nu_3 \rightarrow \nu_2 + a$.} emitting a visible or invisible particle at the one-loop level
\begin{equation}
    \nu_i \rightarrow \nu_j + a \,,
\end{equation}
where $\nu_{i}$ and $\nu_{j}$ are the mass eigenstates and $a$ is the emitted particle. Many studies on the visible or invisible decay of the high-energy neutrinos have been done in the literature \cite{pakvasa2000neutrinos, beacom2003, bustamante2017, denton2018, bustamante2020new, abdullahi2020}. In this section, we investigate the implications of invisible neutrino decay to ALPs on the residual gamma-ray spectra of QSO B1240+326 at sub-PeV energies and its contribution to the cumulative flux from all the FSRQs.
 
We assume normal mass ordering, i.e., $m_1$ $<$ $m_2$ $<$ $m_3$, with the lightest neutrino, $\nu_1$, to be massless and hence stable. Using the current three-flavor neutrino oscillation data from Ref. \cite{esteban2020fate}, we assume $m_2 \approx 8.61$ meV and $m_3 \approx 50.1$ meV in our analysis. While propagating over the cosmological distances, neutrinos will decay into ALPs such that their number $N_i(z)$, with mass eigenstate $\nu_i$, changes with the redshift $z$. The survival probability of neutrinos can be obtained as \cite{huang2023}
\begin{equation}
   \frac{N_i(z)}{N_i(z_0)} = \text{exp}\left(\frac{-m_i}{\tau_iE_{\nu}}D_{eff}(z)\right) \,,
\end{equation}
with $D_{eff}(z)$ as the effective distance given by
\begin{equation}
    D_{eff}(z) = \frac{c}{H_0}\int^{z_0}_{z}\frac{dz^{'}}{(1+z^{'})^2} \frac{1}{\sqrt{\Omega_m (1+z^{'})^3+\Omega_{\Lambda}}} \,.
\end{equation}
Here, $z_0$ is the redshift of the source, $\tau$ is the neutrino decay lifetime, $H_0$ is the Hubble constant, $\Omega_m \approx 0.315$, and $\Omega_{\Lambda} \approx 0.685$.

The total ALP flux arising from $\nu_i$ decays is given by 
\begin{equation}
    \phi_a (E_{\nu}) =  \sum_{\alpha=\mu,e} \mathcal{P}_{\nu_{\alpha}a}(E_{\nu})\,\phi_{\nu_{\alpha}} \label{eq:nudecalp}
\end{equation}
where
\begin{equation}
    \mathcal{P}_{\nu_{\alpha}a}(E_{\nu}) = \sum_{i=2,3}\left[1 - \text{exp}\left(\frac{-m_i}{\tau_iE_{\nu}}D_{eff}(z)\right)\right]|U_{\alpha i}|^2 \,, \label{eq:alpprob}
\end{equation}
is the probability of ALP production from $\nu_{\alpha}$, $\phi_{\nu_{\alpha}}$ is the flux of $\nu_{\alpha}$ at the source, and $U_{\alpha i}$ denotes the leptonic flavor mixing matrix \cite{pdg2022}.
\begin{figure}
    \centering
    \includegraphics[width=\linewidth]{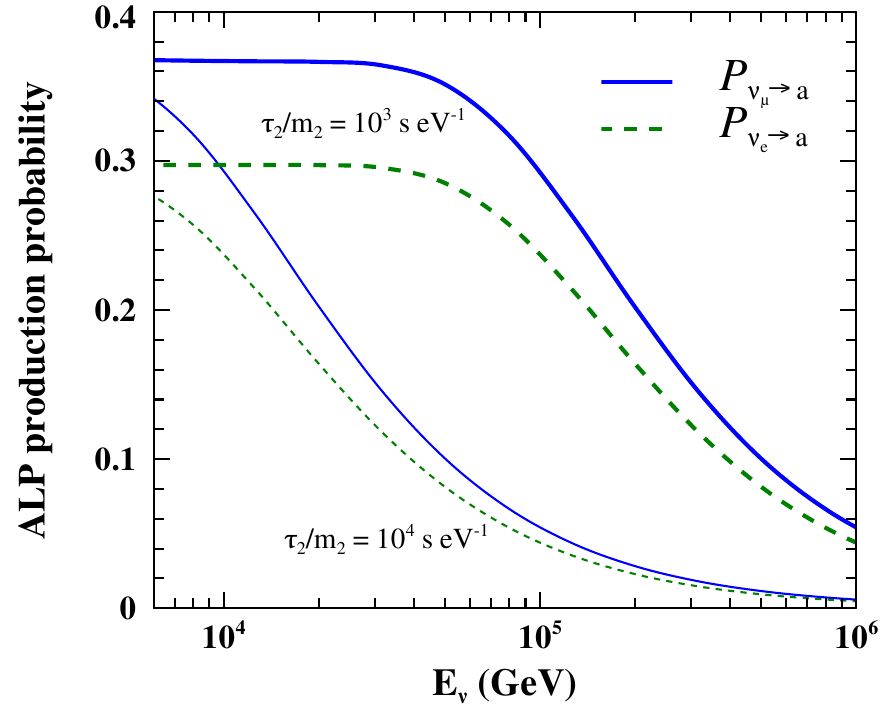}\caption{\label{fig:alpprodprob}ALP production probability from invisible neutrino decay of $\nu_{\mu}$ (solid) and $\nu_{e}$ (dotted) for two benchmark values of $\tau_{2}/m_{2}$.}
\end{figure}

In Eq. \ref{eq:alpprob} above, it can be seen that the ALP production probability depends exponentially on the ratio $\tau_{i}/m_{i}$. Therefore, it is essential to examine the existing bounds on neutrino lifetime. In the literature, several constraints on neutrino lifetime have been proposed \cite{frieman1988neutrino, berryman2015, gonzalez2008, porto2020constraining, aiello2023probing}; we consider the cosmological constraint, $\tau_{\nu}=4\times10^{5} (m_{\nu}/50 \,\text{meV})^5$ s, from Ref. \cite{barenboim2021}. In this work, we assume $\tau_{3}/m_{3} = 10 ^{7}$ s eV$^{-1}$ and two benchmark values of $10 ^{3}$ s eV$^{-1}$ and $10^{4}$ seV$^{-1}$ for $\tau_{2}/m_{2}$. As we can see in Fig. \ref{fig:alpprodprob}, for $\tau_{2}/m_{2} = 10 ^{3}$ s eV$^{-1}$, the ALP production probability is significant ($\sim10\%$) even up to PeV energies, whereas for $\tau_{2}/m_{2} = 10 ^{4}$ s eV$^{-1}$ the probability goes down to below $\sim 3\%$ for energies above 100 TeV.

Using Eq. \ref{eq:nudecalp}, we can then calculate the ALPs flux produced at the source. These ALPs then travel through the intergalactic medium and back-convert into photons upon entering into the Galactic magnetic field and may be observable. In the left panel of Fig. \ref{fig:expgaflux}, we show the contribution from invisible neutrino decay to the gamma-ray flux at sub-PeV energies. We find that for neutrino lifetime $\tau_{2}/m_{2} = 10^{3}$ s eV$^{-1}$, LHAASO-KM2A provides a good observational sensitivity. We also compute the contribution from neutrino decay to diffuse gamma-ray flux as shown in the right panel of Fig. \ref{fig:expgaflux}. We find that although the order of gamma-ray flux is significantly higher compared to counterpart diffuse $\gamma$-rays, it is still negligible to provide any contribution to diffuse $\gamma$ flux from the Galactic plane. Next-generation neutrino detectors like IceCube-Gen2, KM3NeT, and Hyper-Kamiokande will provide more stringent bounds on neutrino decay lifetime. This will open a new window for future $\gamma$-ray studies to search for their footprints and narrow down the hunt of particles beyond the Standard Model.

\begin{acknowledgments}
The author would like to thank the anonymous referee for constructive comments which helped in improving the manuscript.
\end{acknowledgments}
\bibliography{references}

\end{document}